\documentclass[a4paper]{article}

\usepackage{INTERSPEECH2022}
\usepackage{comment}
\usepackage{url}
\usepackage{hyperref}
\usepackage{amsmath,amssymb}
\usepackage{subfigure}
\hypersetup{
    colorlinks = false,
}
\usepackage{cite}

\title{SelfRemaster: Self-Supervised Speech Restoration with \\ Analysis-by-Synthesis Approach Using Channel Modeling}
\name{Takaaki Saeki, Shinnosuke Takamichi, Tomohiko Nakamura, Naoko Tanji, and Hiroshi Saruwatari}
\address{Graduate School of Information Science and Technology, The University of Tokyo, Japan.}
\email{\{takaaki\_saeki, shinnosuke\_takamichi\}@ipc.i.u-tokyo.ac.jp}

\begin{document}

\maketitle
\begin{abstract}\vspace{-1mm}
We present a self-supervised speech restoration method without paired speech corpora.
Because the previous general speech restoration method uses artificial paired data created by applying various distortions to high-quality speech corpora, it cannot sufficiently represent acoustic distortions of real data, limiting the applicability.
Our model consists of analysis, synthesis, and channel modules that simulate the recording process of degraded speech and is trained with real degraded speech data in a self-supervised manner.
The analysis module extracts distortionless speech features and distortion features from degraded speech, while the synthesis module synthesizes the restored speech waveform, and the channel module adds distortions to the speech waveform.
Our model also enables audio effect transfer, in which only acoustic distortions are extracted from degraded speech and added to arbitrary high-quality audio.
Experimental evaluations with both simulated and real data show that our method achieves significantly higher-quality speech restoration than the previous supervised method, suggesting its applicability to real degraded speech materials. 
\end{abstract}

\vspace{-1mm}
\section{Introduction}\vspace{-1mm}
Speech data used in modern speech processing research is usually recorded with sophisticated digital recording equipment and in a well-designed environment.
However, most existing audio data contains acoustic distortions and have different data properties from high-quality audio data.
To expand the range of applications of speech processing, it is necessary to make use of speech data in varying qualities, e.g., speech material recorded with poor or outdated audio devices~\cite{switchboard225858,hansen18_interspeech} and endangered languages' voices~\cite{abraham-etal-2020-crowdsourcing,takamichi-saruwatari-2018-cpjd,Matsuura2020SpeechCO}.
However, restoring or analyzing such degraded audio is challenging because paired high-quality and degraded speech data is not generally available.
The previous general speech restoration method~\cite{liu2021voicefixer} relies on artificial paired corpora for supervised learning.
It cannot sufficiently represent acoustic distortions of real data, which limits the applicability.

We propose a self-supervised speech restoration method without paired speech data.
Our proposed model consists of \textit{analysis}, \textit{synthesis}, and \textit{channel} modules based on the recording process of degraded speech.
The analysis module extracts {\it time-variant} distortionless speech features and {\it time-invariant} acoustic distortion features from degraded speech (hereafter, ``channel feature'').
The other modules carry out the generation process of distorted speech; the synthesis module synthesizes distortionless speech (i.e., restored speech), and the channel module adds acoustic distortion.
The model is trained in a self-supervised manner by minimizing the reconstruction loss between degraded input and reconstructed speech.
We also propose a dual-learning method for stable self-supervised learning, where a backward process defines distortionless speech feature loss with arbitrary high-quality speech corpora while the forward process calculates the reconstruction loss.
Experimental evaluations show that our method achieves significantly higher-quality speech restoration than the previous supervised method.
Our implementation\footnote{\scriptsize{\url{https://github.com/Takaaki-Saeki/ssl\_speech\_restoration}}} and audio samples\footnote{\scriptsize{\url{https://takaaki-saeki.github.io/ssl\_remaster\_demo/}}} are publicly available.
The contributions of this paper are as follows:
\vspace{-1mm}
\begin{itemize} \leftskip -5.5mm \itemsep -0.5mm
    \item We propose self-supervised speech restoration that can learn various acoustic distortions without a paired speech corpora and is applicable to real degraded speech data.
    \item Our method achieved significantly higher-quality speech restoration than the previous supervised method on both simulated data and real historical materials.
    \item Our model enables audio effect transfer, in which only channel features are extracted from degraded speech and applied to other high-quality speech.
\end{itemize}

\vspace{-1mm}
\section{Related work}\vspace{-1mm}
\textbf{Speech restoration}. Many previous studies have focused on specific speech restoration tasks such as bandwidth extension~\cite{bandwidth4100688,Kuleshov2017AudioSR}, dereverberation~\cite{derev6963349,Ernst2018SpeechDU}, denoising~\cite{Luo2019ConvTasNetSI,Koizumi2020SpeechEU}, and declipping~\cite{Bie2015DetectionAR,Zvika2021ASA}.
Similar to our work, some previous studies~\cite{Han15derev,Shu2021JointEC} jointly perform multiple speech restoration tasks, and some methods leverage waveform synthesis~\cite{Polyak2021HighFS,liu2021voicefixer} to perform speech restoration.
Unlike these studies, our work does not focus on removing additive noises and enables self-training.
A previous study on general speech restoration with a neural vocoder~\cite{liu2021voicefixer} is the closest work to ours.
The most significant difference is that our method does not require paired data, so it can be trained with real audio materials in a self-supervised manner instead of artificially creating degraded speech data.
Furthermore, as our proposed method also extracts channel features at the same time, it can transfer audio effects, which adds the desired channel distortion to arbitrary audio.

\textbf{Self-supervised learning for speech}. Self-supervised speech representation learning~\cite{baevski2020wav2vec,hsu2021hubert,chen2022wavlm} and methods using those models for various downstream tasks~\cite{Yi2020ApplyingWT,Nguyen2020InvestigatingSP} have widely been studied.
There are several studies related to our work which involve self-supervised learning on speech enhancement~\cite{Wang2020SelfsupervisedLF,zezario20self,Du2020SelfSupervisedAM}, though they only focus on denoising while our method restores speech from various kinds of distorted speech.
Similar to our method, the DDSP autoencoder~\cite{Engel2020DDSP} learns disentangled features from acoustic signal in a self-supervised manner and manipulates each feature.
It uses a simple sinusoidal vocoder and assumes only reverberation as acoustic distortion, while our method leverages a more expressive waveform synthesis model to achieve high-quality speech restoration and channel modules to capture various acoustic distortions.

\vspace{-1mm}
\section{Proposed method}

\subsection{Basic framework}\label{sec:method-basic}\vspace{-1mm}

\begin{figure}[t]
  \centering
  \includegraphics[width=1.0\linewidth, clip]{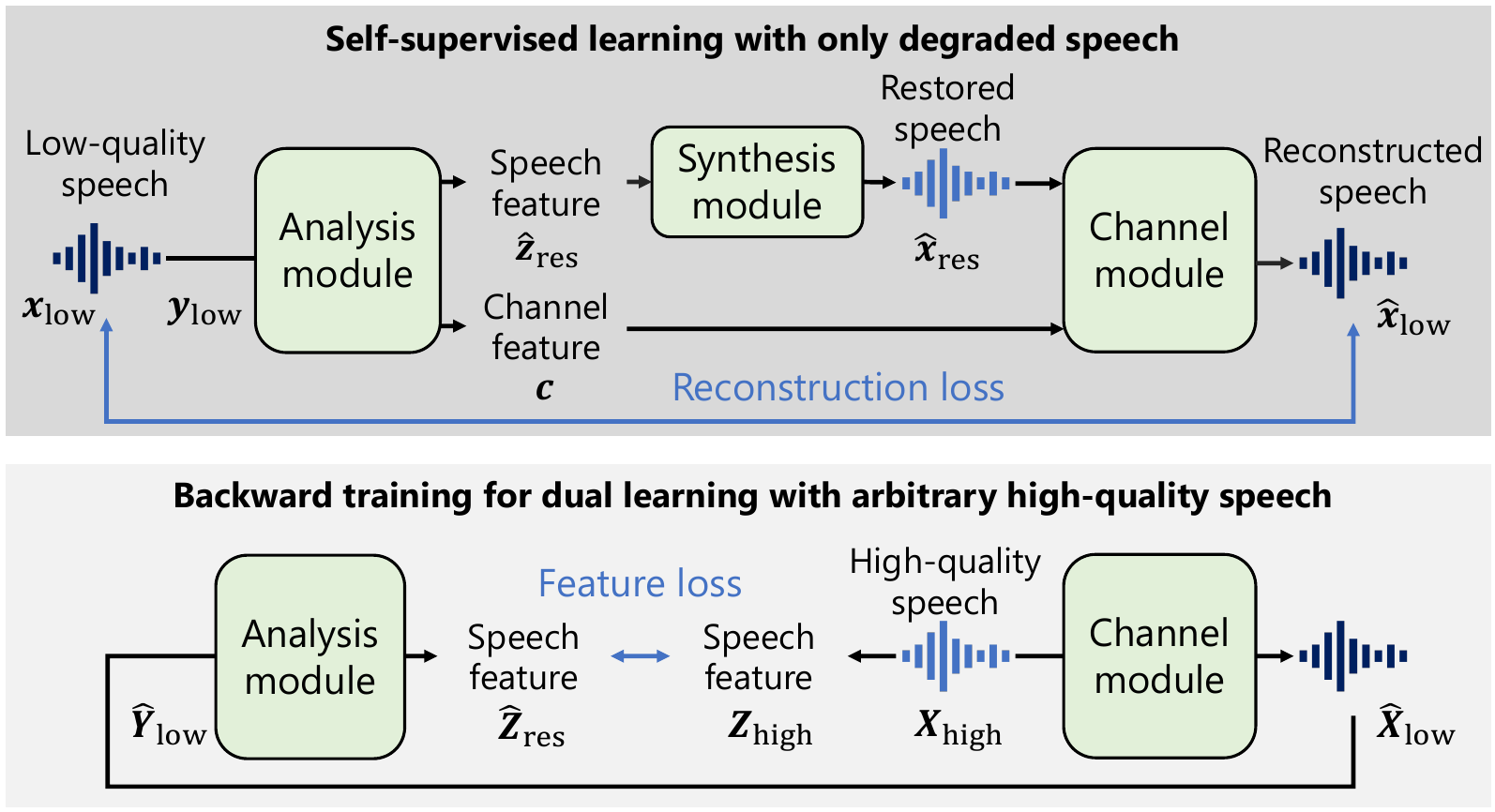}
  \caption{Proposed self-supervised speech restoration method, consisting of analysis, synthesis, and channel modules that simulate recording process of degraded speech. Basic training process (top) minimizes reconstruction loss between degraded and reconstructed speech. We also propose dual learning (top \& bottom) with arbitrary high-quality speech corpora.}
  \label{fig:dual-learning}
  \vspace{-4mm}
\end{figure}

We begin with an overview of the recording process of degraded speech underlying our method.
High-quality speech (i.e., distortionless speech) is emitted from the mouth through a human speech production process, which can be parameterized by {\it time-variant} speech features such as traditional source-filter vocoder features or mel spectrograms.
Then acoustic distortions (e.g., non-linear response of recording equipment and lossy audio coding) are added to the high-quality speech, resulting in a final recorded audio.
We assume the distortions (i.e., channel features) to be {\it time-invariant}, where the recording equipment or audio coding does not change in each audio sample.

The proposed speech restoration model consists of analysis, synthesis, and channel modules based on the abovementioned process, as shown in Fig.~\ref{fig:dual-learning}~(top).
All of these modules are composed of neural networks, so they can be trained in an end-to-end manner.
We denote the degraded-speech waveform and the degraded-speech features as $\Vec{x}_{\mathrm{low}}$ and $\Vec{y}_{\mathrm{low}}$, respectively.
The analysis module estimates the speech features of the restored speech, $\hat{\Vec{z}}_{\mathrm{res}}$, from $\Vec{y}_{\mathrm{low}}$.
At the same time, the intermediate features of the analysis module are input to other layers in order to finally estimate the time-invariant channel features $\Vec{c}$.
Note that we use mel spectrogram for $\Vec{y}_{\mathrm{low}}$ and the two features described in Section~\ref{sec:evaluation-condition} for $\hat{\Vec{z}}_{\mathrm{res}}$.
The synthesis module simulates human speech production; it takes $\hat{\Vec{z}}_{\mathrm{res}}$ as input and estimates the speech waveform of the restored speech $\hat{\Vec{x}}_{\mathrm{res}}$.
We use HiFi-GAN~\cite{Kong2020HiFiGANGA} for the synthesis module, where it is trained on an arbitrary high-quality speech corpus and the model parameters are frozen.
The channel module simulates the acoustic distortion; it is conditioned by the channel feature $\Vec{c}$ and distorts the restored speech $\hat{\Vec{x}}_{\mathrm{res}}$ to estimate the degraded speech $\hat{\Vec{x}}_{\mathrm{low}}$.
The model is trained to minimize the frame-level reconstruction loss, which can be defined as a multi-scale spectral loss~\cite{Engel2020DDSP} as follows:
\begin{equation}\label{eq:recons}
    \mathcal{L}_{\mathrm{recons}} = \sum_{i} \{ ||\Vec{s}_{i} - \hat{\Vec{s}}_{i}||_{1} + \alpha ||\log \Vec{s}_{i} - \log \hat{\Vec{s}}_{i}||_{1} \},
\end{equation}
where $\Vec{s}_{i}$ and $\hat{\Vec{s}}_{i}$ are the amplitude spectrograms of $\Vec{x}_{\mathrm{low}}$ and $\hat{\Vec{x}}_{\mathrm{low}}$, respectively.
The subscript $i$ is the window length of the Fourier transform and $\alpha$ is a weight of the log term, where we used $i = (2048, 1024, 512, 256, 128, 64)$ and $\alpha=1.0$ in our experiment.
Speech restoration is performed by driving the trained analysis and synthesis modules. The analysis module receives the acoustic feature of degraded speech, $\Vec{y}_{\mathrm{low}}$, from the analysis module, and the synthesis module synthesizes the restored-speech waveform $\hat{\Vec{x}}_{\mathrm{res}}$.

Furthermore, the trained model can also perform an ``audio effect transfer'' in addition to speech restoration.
The audio effect is transferred without the synthesis module. We first extract the channel feature $\Vec{c}$ from the degraded speech $\Vec{x}_{\mathrm{low}}$ using the trained analysis module. The trained channel module conditioned by $\Vec{c}$ receives arbitrary high-quality audio and distorts it so that the resulting audio sounds distorted like the referred degraded speech. 

\begin{figure}[t]
  \centering
  \includegraphics[width=0.82\linewidth, clip]{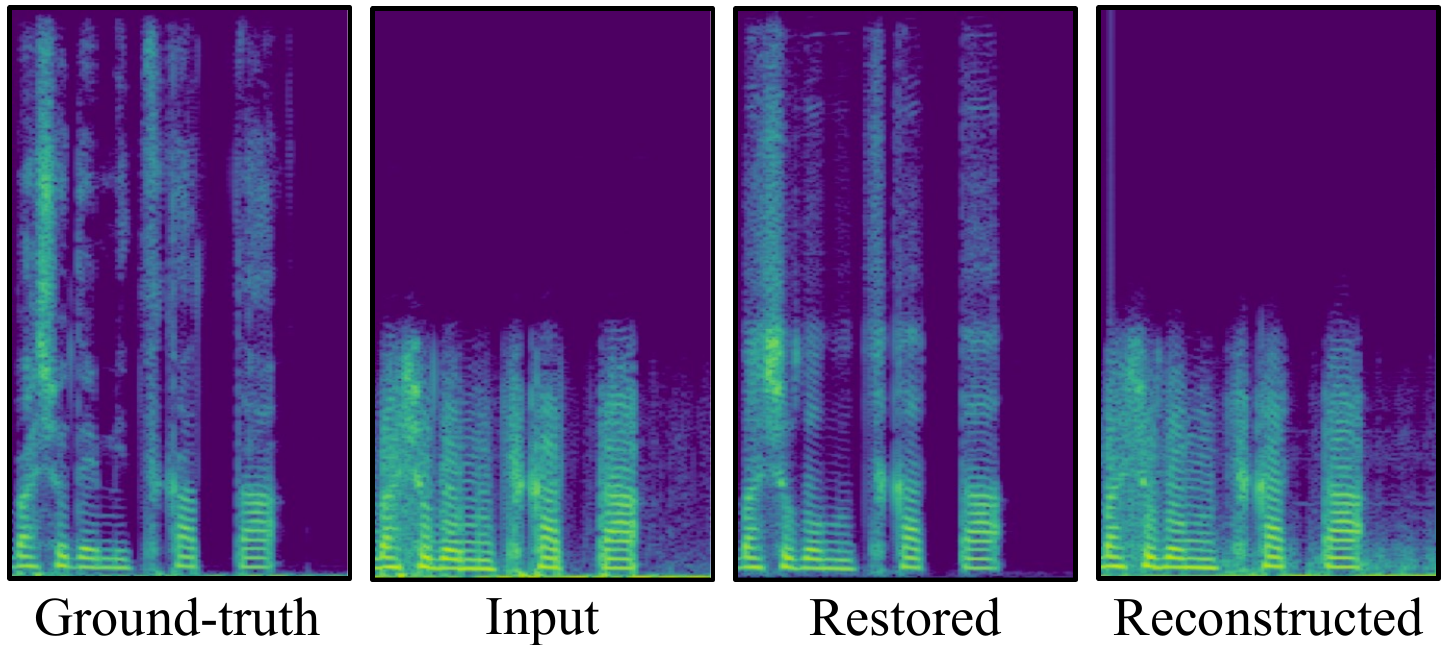}
  \vspace{-1mm}
  \caption{Spectrograms obtained with proposed method}
  \label{fig:specs-ssl}
  \vspace{-5mm}
\end{figure}

Fig.~\ref{fig:specs-ssl} shows the spectrograms obtained by the proposed method.
The input low-quality speech is simulated on the basis of the quantization and resampling described in Section~\ref{sec:evaluation}, and is missing high-frequency bands that are present in the ground-truth high-quality speech and has distorted low-frequency bands.
In the restored speech, the missing and distorted bands are restored.
Furthermore, the reconstructed speech output by the channel module faithfully reproduces the input speech.

Unfortunately, the above basic self-supervised training process is unstable because each module has a high expressive power, and the modules are only trained with the reconstruction loss between the degraded and reconstructed waveform.
Specifically, the analysis module can represent the effect of the channel module, so there is no guarantee that the analysis module outputs the speech features of high-quality speech.
Therefore, we propose the dual-learning method described in Section~\ref{sec:method-dual}.
In Section~\ref{sec:method-pre}, we also introduce supervised pre-training to improve the performance when only a very limited amount of data is available for the model size, such as when we use rare historical audio.

\vspace{-1mm}
\subsection{Dual learning for stable self-supervised learning}\label{sec:method-dual}\vspace{-1mm}
Fig.~\ref{fig:dual-learning} shows the proposed dual-learning method.
In addition to the basic training framework described in Section~\ref{sec:method-basic}, we introduce a training task that propagates information in the backward direction.
We denote the two training tasks in the forward direction and backward direction as $\mathcal{T}_{\mathrm{forward}}$ and $\mathcal{T}_{\mathrm{backward}}$, respectively.
$\mathcal{T}_{\mathrm{forward}}$ is the basic learning framework described in Section~\ref{sec:method-basic}; degraded speech is sent to the analysis module to reconstruct the degraded waveform.
Meanwhile, in $\mathcal{T}_{\mathrm{backward}}$, high-quality speech is input to the channel module to estimate the features of high-quality speech.
Our method is analogous to the forward and backward learning between two domains in machine translation~\cite{hedualmachine2016}.

Here, we use arbitrary high-quality speech, which does not align with the degraded speech and only consists of other speaker's utterances. 
The high-quality speech waveform $\Vec{X}_{\mathrm{high}}$ is first passed through the channel module to generate the degraded speech waveform $\hat{\Vec{X}}_{\mathrm{low}}$ with channel features.
Next, the speech features of the degraded speech, $\hat{\Vec{Y}}_{\mathrm{low}}$, are input to the analysis module to estimate the speech feature of the restored speech, $\hat{\Vec{Z}}_{\mathrm{res}}$.
Then a feature loss is defined as the mean squared error between $\hat{\Vec{Z}}_{\mathrm{res}}$ and the speech feature $\Vec{Z}_{\mathrm{high}}$ is obtained from the high-quality speech as:
\begin{equation}
    \mathcal{L}_{\mathrm{feature}} = ||\Vec{Z}_{\mathrm{high}} - \hat{\Vec{Z}}_{\mathrm{res}}||_{2}.
\end{equation}
The final loss function is given as the weighted sum of the loss function between the reconstruction loss and the feature loss as:
\begin{equation}\label{eq:full}
    \mathcal{L} = (1 - \beta) * \mathcal{L}_{\mathrm{recons}} + \beta * \mathcal{L}_{\mathrm{feature}},
\end{equation}
where $\beta$ is a weight of the feature loss.
Here, the gradient due to feature loss is not propagated to the channel module, and the channel module is only trained on the reconstruction loss.
Intuitively, $\mathcal{T}_{\mathrm{forward}}$ and $\mathcal{T}_{\mathrm{backward}}$ can be interpreted as:
\begin{itemize}\leftskip -5.5mm
    \item $\mathcal{T}_{\mathrm{forward}}$ obtains the analysis module that adapts to degraded speech and the channel module that expresses the acoustic distortion.
    \item $\mathcal{T}_{\mathrm{backward}}$ obtains the analysis module to estimate high-quality speech features.
\end{itemize}
This dual learning method enables self-supervised learning of the analysis module and channel module without paired data.

\vspace{-1mm}
\subsection{Supervised pretraining for low-resource settings}\label{sec:method-pre}\vspace{-1mm}
Data scarcity is an obstacle in restoring historical audio materials; only a small amount of audio is available due to the limited amount of recording storage size available in the past.
Thus, we introduce a supervised pre-training method to obtain the initial weights for such low data resource conditions as:
\begin{enumerate}\leftskip -5.5mm
    \item We supervisedly train the analysis and channel modules using paired data artificially created from arbitrary high-quality speech data.
    \item With the initialization from supervised pre-training, we then carry out the self-supervised learning described in Section~\ref{sec:method-basic} and Section~\ref{sec:method-dual} with real degraded speech data.
\end{enumerate}
In the supervised pretraining, we generate pseudo-degraded speech waveform $\Vec{X}_{\mathrm{pseudo-low}}$ by randomly applying distortion to the original high-quality speech.
When we denote the restored speech feature estimated from the pseudo-degraded speech feature $\Vec{Y}_{\mathrm{pseudo-low}}$ as $\hat{\Vec{Z}}_{\mathrm{res}}$, we define the feature loss as $\mathcal{L}_{\mathrm{feature}} = ||\Vec{Z}_{\mathrm{high}} - \hat{\Vec{Z}}_{\mathrm{res}}||_{2}$.
Note that we input the high-quality speech waveform to the channel module instead of using the output of the synthesis module.
Then we train the model to minimize the weighted sum of the reconstruction loss and feature loss in the same manner as Eq.~\eqref{eq:full}.

\vspace{-1mm}
\section{Experimental evaluations}\label{sec:evaluation}

\subsection{Experimental condition}\label{sec:evaluation-condition}\vspace{-1mm}

\textbf{Database.}
We used both simulated and real degraded speech datasets with the sampling rate set to 22.05~kHz.
We created simulated datasets by applying various distortions to the JSUT corpus~\cite{jvsjsut20}, which consists of around six hours of high-quality speech utterances of a Japanese female speaker.
We randomly selected 25 and 25 sentences for the validation and test sets, respectively. 
For the real data, we used a Japanese historical speech dataset~\cite{tono}.
This dataset was recorded in the 1960s--1970s and consists of nine speakers telling folktales, and the whole duration was about 22-minutes.
The analog audio clips recorded on the cassette tape were digitized using a radio cassette player.
Because this data contains a lot of additive noise, iZotope RX9 was used for noise reduction as preprocessing.
We randomly selected two-minute audio clips for the validation and test sets.
We used the JVS corpus~\cite{jvsjsut20} as a high-quality speech corpus for pretraining the synthesis module, supervised pretraining in Section~\ref{sec:method-pre}, and $\mathcal{T}_{\mathrm{backward}}$ of dual learning.

\textbf{Speech feature analysis.}
We used two types of speech features $\hat{\Vec{z}}_{\mathrm{res}}$: mel spectrogram and source-filter vocoder features. The 80-dimensional mel spectrogram was extracted with a frame size of 1024 and frame shift of 256. The vocoder features consists of F0 and 41-dimensional mel cepstrum coefficients with a 5-ms frame shift, which were extracted by the WORLD vocoder~\cite{morise16world}. Hereafter, we denote these types as ``Melspec'' and ``SourceFilter'', respectively.

\begin{table*}[tb]
\centering
\caption{Evaluation results with simulated data}
\vspace{-2mm}
\label{tab:eval-simulation}
\scalebox{0.9}{
\begin{tabular}{l| cc cc cc cc}
\toprule
                        & \multicolumn{2}{c}{(a) Band-limited} & \multicolumn{2}{c}{(b) Clipped} & \multicolumn{2}{c}{(c) Quantized \& Resampled} & \multicolumn{2}{c}{(d) Overdrive} \\ \cmidrule(lr){2-3} \cmidrule(lr){4-5} \cmidrule(lr){6-7} \cmidrule(lr){8-9}
                        & MCD               & MOS               & MCD             & MOS            & MCD                    & MOS                    & MCD               & MOS              \\ \midrule
Ground-truth            & - & $4.51 \pm 0.092$  &   - & $4.58 \pm 0.083$ &  - &  $4.67 \pm 0.098$   &  - & $4.65 \pm 0.098$  \\
Input                   & $17.85$ & $2.38 \pm 0.115$ &   $11.21$ & $2.45 \pm 0.124$  &    $22.25$ & $1.73 \pm 0.115$   &  $16.73$ & $1.54 \pm 0.105$ \\
Liu, et al.~\cite{liu2021voicefixer} &  $6.98$  & $3.74 \pm 0.115$   &  $9.33$  & $3.01 \pm 0.119$  &  $9.34$  & $2.80 \pm 0.109$   &   $10.23$  & $2.00 \pm 0.112$ \\
SSL-dual (MelSpec)      &  $\textbf{5.61}$ & $\textbf{4.20} \pm 0.090$  & \textbf{7.20} & $\textbf{3.49} \pm 0.114$  &  $\textbf{8.07}$ & $\textbf{3.27} \pm 0.111$      &   $\textbf{8.93}$ & $\textbf{2.68} \pm 0.108$    \\
SSL-dual (SourceFilter) &   $5.88$ &  $3.46 \pm 0.115$   &  $7.59$ & $2.49 \pm 0.190$  &  $9.60$ & $2.66 \pm 0.109$   &  $9.06$ &  $2.58 \pm 0.125$    \\ \bottomrule
\end{tabular}
}
\vspace{-3mm}
\end{table*}

\textbf{Model details.}
We used the U-Net architecture~\cite{Ronneberger2015UNetCN} for the analysis and channel modules.
In each down sampling of the U-Net, the temporal resolution was reduced by half with four layers of residual convolution blocks and average pooling.
Each residual convolution block consists of convolutional layers and batch normalization layers~\cite{iofee15batchnorm} with a skip connection~\cite{He2016DeepRL}.
During up-sampling, the temporal resolution is doubled by applying deconvolution, resulting in time-variant high-quality speech features with the same temporal resolution as that of the input features.
For more details, refer to the published implementation.
For the synthesis module, we used the pretrained multi-speaker model\footnote{\scriptsize{\url{https://github.com/jik876/hifi-gan}}} under the ``MelSpec'' condition and we trained the model with the JVS corpus under the ``SourceFilter'' condition. 

\textbf{Training settings.}
The batch size was four and the maximum number of epochs was set to 50, and the model used for evaluation was selected based on the validation results.
Adam~\cite{kingma14adam} was used as the optimizer with the initial learning rate set to $0.001$.
We applied a learning rate scheduling that multiplies the learning rate by 0.5 if the validation loss did not decrease over three epochs.
In dual learning and supervised pretraining, we set $\beta$ of Eq.~\eqref{eq:full} to $0.1$ and $0.001$, respectively.
For the evaluation with real data, we used the supervised pretraining described in Section~\ref{sec:method-pre}, in which quantization and resampling were randomly applied to each speech waveform.
After $\mu$-law quantization~\cite{recommendation1988pulse} with random quantization bits ranging from 6 to 10 bits, the data was resampled with random sampling frequencies of \{8, 11.25, 12, 16\}~kHz.

\vspace{-1mm}
\subsection{Evaluation on speech restoration with simulated data}\label{sec:eval-simu}\vspace{-1mm}

We evaluated the proposed method by adding various artificial distortions to the high-quality speech corpus described in Section~\ref{sec:evaluation-condition}.
The purpose of this evaluation is to investigate the performance in an experimental setting similar to that of the previous supervised speech restoration model~\cite{liu2021voicefixer}.
We compared our proposed method described in Section~\ref{sec:method-dual} (``SSL-dual'') with the previous supervised model (``Liu, et al.'') for generalized speech restoration tasks~\cite{liu2021voicefixer}, where we used a pretrained model provided by the author\footnote{\scriptsize{\url{https://github.com/haoheliu/voicefixer}}}.
This supervised model is trained with artificial paired data generated by randomly applying noise, reverberation, clipping, and low-pass filtering to multi-speaker and multi-lingual clean speech corpora.

We created datasets simulating two in-domain and two out-of-domain distortions. ``In-domain'' and ``out-of-domain'' indicate whether the training data of the supervised model includes the distortion or not. The in-domain distortions are
(a) \textbf{Band-limited}: Applying biquad-lowpass filtering to the high-quality speech with a cut-off frequency of 4~kHz;
and (b) \textbf{Clipped}: Clipping the high-quality speech waveform with an absolute-value threshold of 0.25. 
The out-domain ones are more significant distortions than in-domain ones, which are
(c) \textbf{Quantized \& resampled}: Quantizing the high-quality speech to 8-bit $\mu$-law and resampling it to $8$~kHz;
and (d) \textbf{Overdrive}: Overdriving the high-quality speech using the SoX implementation~\cite{sox15}.

We conducted an objective evaluation using mel cepstral distortion (MCD)~\cite{fukada92melcep} with the dimension of the cepstrum set to 24.
We also evaluated with subjective mean opinion score (MOS) test, where we recruited 40 Japanese speakers for each evaluation and calculated MOS with 95~\% confidence intervals.
Table~\ref{tab:eval-simulation} lists the results, with the best results shown in bold.

As shown in the table, ``SSL-dual (Melspec)'' significantly improves the quality of input degraded speech in all cases, indicating that it can greatly improve the speech quality.
We also verified that the proposed method significantly outperforms the previous supervised method in both out-of-domain and in-domain cases.
Particularly, in the simple band-limited case, the proposed method achieves a high MOS of $4.20$.
In the proposed methods, the quality of ``Melspec'' is higher than that of ``SourceFilter''.
The difference in MOS may be due to the lower naturalness of the neural vocoder in the SourceFilter, while MCD shows a significant improvement over the previous supervised model.

\vspace{-1mm}
\subsection{Evaluation on speech restoration with real data}\label{sec:eval-real}\vspace{-1mm}
We also evaluated the proposed method with real data.
Note that there is no ground-truth speech in this data.
Because this is a small dataset, we applied supervised pre-training as shown in Section~\ref{sec:evaluation-condition}.
Table~\ref{tab:result_real-data}(a) lists the results of the MOS test.
The suffix ``-pre'' indicates the use of the supervised pre-training.

The proposed method ``SSL--dual--pre (Melspec),'' achieved the highest score and significantly outperformed the previous supervised model.
Here, the performance of the previous method and some of the proposed methods were lower than the input, which is most likely due to the quality degradation in the neural vocoder.
We also determined that the MOS of both MelSpec and SourceFilter can be improved by supervised pre-training.
``SSL--dual--pre (Melspec)'' showed a higher MOS than the input but there was not significant difference.
Therefore, we additionally conducted an AB test with 40 Japanese evaluators to clarify the difference in quality.
Table~\ref{tab:result_real-data}(b) indicate that the proposed method achieved significantly higher preference score than the input.
The above evaluation results show that our proposed method is feasible when only a small amount of real data is available.

\begin{table}[tb]
    \centering
    \caption{Evaluation results with real data}
    \label{tab:result_real-data}
    \vspace{-3mm}
    \subtable[\textbf{MOS tests of all methods}]{
        \vspace{-4mm}
        \footnotesize
        \begin{tabular}{l|c}
        \toprule
                            &  MOS    \\ \midrule
        Input               &   $2.98 \pm 0.136$   \\ \midrule
        Liu, et al.~\cite{liu2021voicefixer}    &   $2.80 \pm 0.138$   \\ 
        SSL--pre (MelSpec)  &   $2.70 \pm 0.127$   \\
        SSL--dual (MelSpec)    &   $2.96 \pm 0.135$   \\
        SSL--dual--pre (MelSpec) &  $\textbf{3.06} \pm 0.140$   \\
        SSL--pre (SourceFilter)  &   $2.14 \pm 0.123$   \\
        SSL--dual (SourceFilter)    &   $2.41 \pm 0.124$   \\
        SSL--dual--pre (SourceFilter)  &  $2.55 \pm 0.135$   \\ \bottomrule
        \end{tabular}
    }
    \subtable[\textbf{Preference AB test}]{
        \vspace{-3mm}
        \footnotesize
        \begin{tabular}{cc}
            \toprule
            SSL--dual--pre (MelSpec) vs. Input   & $p$-value \\ \midrule
            {\bf 0.565} - 0.435 & 0.047 \\ \bottomrule
        \end{tabular}
    }
\end{table}

\begin{table}[tb]
    \centering
    \caption{Evaluation results of audio effect transfer}
     \vspace{-2mm}
    \footnotesize
    \begin{tabular}{l|cc}
    \toprule
                              &  Simulated &  Real    \\ \midrule
    Reference & $3.98 \pm 0.152$ &  $2.99 \pm 0.146$    \\ \midrule
    High-quality & $1.16 \pm 0.069$  & $1.30 \pm 0.093$   \\
    Mean spec. diff. & $1.68 \pm 0.113$  & -    \\ 
    Proposed   & $\textbf{3.44} \pm 0.159$ & $\textbf{2.12} \pm 0.155$ \\ \bottomrule
    \end{tabular}
    \vspace{-4mm}
\end{table}

\vspace{-1mm}
\subsection{Evaluation on audio effect transfer}\vspace{-1mm}
To investigate the effectiveness of the audio effect transfer, we retrieved the acoustic distortion from the training data and added it to different high-quality speech data.
There were two kinds of training data: ``Simulated'' generated with quantization and resampling as in Section~\ref{sec:eval-simu} and ``Real'' used in Section~\ref{sec:eval-real}. The high-quality speech to be modified was sampled from the JVS corpus.
We conducted a five-level similarity MOS (SMOS) test with 40 listeners to evaluate the degree to which the output speech sounded distorted like the training data.
``Reference'' is a sample that has the same channel features as the training data.
For comparison, we prepared a method (``Mean spec. diff'') that multiplies the amplitude spectrum of the high-quality speech by the time-averaged amplitude spectrum difference of training data and high-quality data.

The results show that the proposed method is closer to the target channel features than the original high-quality speech on both simulated and real data.
Regarding the results on the simulated data, the SMOS is significantly higher than that of the case corrected by the mean amplitude spectral difference.
Although the score of ``Proposed'' is 0.5--0.8 smaller than the reference, it performed sufficiently.

\vspace{-1mm}
\section{Conclusion}\vspace{-1mm}
We proposed a self-supervised speech restoration method consisting of analysis, synthesis, and channel modules.
Our method achieved significantly higher-quality speech restoration than the previous supervised method in several simulated data and was effective even with low-resource real data.
Our future work includes applying our method to a wider variety of real data, such as where multiple speakers are speaking at the same time.

\textbf{Acknowledgements:}
Part of this work was supported by JSPS KAKENHI Grant Number 21H04900.

\bibliographystyle{IEEEtran}
\bibliography{tts}

\end{document}